\documentclass[12pt]{iopart}
\usepackage{amsmath}
\usepackage{graphicx}
\usepackage{siunitx}
\usepackage{hyperref}
\usepackage{lineno}
\usepackage{multirow}
\usepackage{makecell}
\usepackage{booktabs} 
\usepackage{textcomp}
\usepackage{bm}
\usepackage[style=ieee,doi=false,url=false]{biblatex}
\bibliography{ref}

\usepackage{iopams}  
\begin{document}

\title[]{Physics-informed 4D X-ray image reconstruction from ultra-sparse spatiotemporal data}


\author{Zisheng Yao$^{1*}$, Yuhe Zhang$^1$, Zhe Hu$^1$, Robert Klöfkorn$^2$, Tobias Ritschel$^3$ and Pablo Villanueva-Perez$^1$}

\address{$^1$ Division of Synchrotron Radiation Research and NanoLund, Lund University, Box 118, 22100, Lund, Sweden}
\address{$^2$ Center for Mathematical Sciences, Lund University, Box 117, 22100, Lund, Sweden}
\address{$^3$ Department of Computer Science, 
University College London, WC1E~6BT London, UK}
\ead{zisheng.yao@sljus.lu.se}
\vspace{10pt}
\begin{indented}
\item[]March 2025
\end{indented}

\begin{abstract}
The unprecedented X-ray flux density provided by modern X-ray sources offers new spatiotemporal possibilities for X-ray imaging of fast dynamic processes. 
Approaches to exploit such possibilities often result in either i) a limited number of projections or spatial information due to limited scanning speed, as in time-resolved tomography, or ii) a limited number of time points, as in stroboscopic imaging, making the reconstruction problem ill-posed and unlikely to be solved by classical reconstruction approaches.
4D reconstruction from such data requires sample priors, which can be included via deep learning (DL). 
State-of-the-art 4D reconstruction methods for X-ray imaging combine the power of AI and the physics of X-ray propagation to tackle the challenge of sparse views. 
However, most approaches do not constrain the physics of the studied process, i.e., a full physical model.
Here we present 4D physics-informed optimized neural implicit X-ray imaging (4D-PIONIX), a novel physics-informed 4D X-ray image reconstruction method combining the full physical model and a state-of-the-art DL-based reconstruction method for 4D X-ray imaging from sparse views.
We demonstrate and evaluate the potential of our approach by retrieving 4D information from ultra-sparse spatiotemporal acquisitions of simulated binary droplet collisions, a relevant fluid dynamic process.
We envision that this work will open new spatiotemporal possibilities for various 4D X-ray imaging modalities, such as time-resolved X-ray tomography and more novel sparse acquisition approaches like X-ray multi-projection imaging, which will pave the way for investigations of various rapid 4D dynamics, such as fluid dynamics and composite testing.  
\end{abstract}
\vspace{2pc}
\noindent{\it Keywords\/}: Ultrafast X-ray Imaging, Physics-informed, Deep learning, Ultra-sparse spatiotemporal data, Four-dimensional (4D) reconstruction

%
%

\section{Introduction}

Over the past few decades, the developments of modern large-scale X-ray facilities (synchrotron light sources and X-ray free-electron lasers) have boosted the field of X-ray imaging. 
Specifically, the enhanced flux density provided by such facilities opens up new possibilities to explore new spatiotemporal resolutions~\cite{wang2008ultrafast,Olbinado2017_MHz_XRI,Vagovic2019_MHzXFEL,yao2024new}, which are crucial for non-destructive 4D studies of fast dynamics under in-situ and operando conditions. 
Such studies have wide application perspectives in various scientific and engineering fields, such as fluid dynamics~\cite{Tekawade2020_fluid}, additive manufacturing~\cite{Makowska2023_AM}, and energy materials~\cite{ziesche20204d}.\\ 

\noindent
In order to fully exploit the unique capabilities of modern X-ray facilities for 4D imaging studies, it is crucial to develop practical 4D reconstruction tools. 
Standard 4D reconstruction for time-resolved tomography, the state-of-the-art 4D imaging technique at modern large-scale X-ray facilities, relies on stacking individually reconstructed 2D slices into a 3D volume at each time point. 
The most established reconstruction methods include analytical reconstruction~\cite{ziegler2007noise}, such as filtered-back projection (FBP), and iterative reconstruction~\cite{ning2014x}, such as simultaneous algebraic reconstruction technique (SART). 
However, the reconstruction quality rapidly degrades when sparse spatiotemporal data are provided, e.g., a limited number of projections or stroboscopic temporal acquisitions~\cite{yu2023reconstruction}.
A practical example is X-ray Multi-projection Imaging (XMPI)~\cite{xmpi2018,duarte2019computed,voegeli2020multibeam}, which is a rotation-free technique capable of capturing faster dynamics than time-resolved tomography at the cost of providing ultra-sparse projection angles of the observed dynamics. 
Under such circumstances, the reconstruction problems become extremely ill-posed~\cite{kabanikhin2011inverse}.
Moreover, when we study the dynamics, the fast motion of the observed process might worsen the recorded images, resulting in fewer usable time points for retrieving the 4D dynamics~\cite{xmpiXFEL}. 
The aforementioned challenges highlight the limitations of current 2D reconstruction approaches to address a 4D problem.
Therefore, we desire a workflow that directly reconstructs 4D dynamics. 
Such a workflow should take into account 4D priors, such as the shape and the motion of the observed object~\cite{zhang2013iterative,engelmann2017samp}.\\

\noindent
Thanks to recent advances in artificial intelligence (AI) and deep learning (DL), capturing complex sample priors~\cite{ulyanov2018deep} becomes applicable, paving the way for image reconstruction from sparse spatiotemporal data.
Specifically, neural radiance fields (NeRF)~\cite{mildenhall2021nerf,yu2021pixelnerf} based methods open new opportunities for sparse-view reconstruction in various 3D and 4D imaging modalities. 
Although the NeRF-based methods originate from visible light setup, they have been adapted to X-ray imaging by embedding the law of X-ray propagation~\cite{zhang2023onix,cai2024structure,corona2022mednerf,zheng2023ultrasparse} and some of them are capable of 4D reconstruction~\cite{zhang20244d}. 
However, such X-ray methods typically require large datasets for training, either by data augmentation~\cite{zheng2023ultrasparse} or by reproducing similar X-ray experiments~\cite{zhang20244d}. 
In essence, such methods provide proper 4D reconstructions benefiting from large datasets and the generalization ability provided by data-driven DL methods, failing to tackle the challenges given by sparsity in the temporal domain.
In the meantime, the full physical model of the studied dynamics can act as an important sample prior for the 4D reconstruction task. 
Specifically, physics-informed neural network (PINN)~\cite{raissi2019physics,pinn2021,cuomo2022scientific} is an emerging tool that can boost the performance of 4D reconstruction. 
By applying the full physical model to the reconstruction workflow, super-resolution~\cite{shone2023deep,yang2024super} in the temporal domain becomes achievable.
In essence, a combination of the NeRF-based method and the physics-informed method~\cite{chu2022physics,duan2024physics,maul2024physics} can potentially address the difficulties caused by ultra-sparse spatiotemporal acquisitions. 
However, such a combination in the field of ultra-fast 4D X-ray imaging still requires investigation.\\

\noindent
In this work, we present 4D-PIONIX, a novel physics-informed 4D X-ray image reconstruction method to tackle the challenges from both limited views and limited time points.
For validation of our proposed method, we simulate the 4D droplet collision process based on the experimental setup of XMPI~\cite{xmpiXFEL} and solve the 4D reconstruction problem from ultra-sparse spatiotemporal data, e.g., two 23.8-degree-apart projections at a limited number of time points. 
The results indicate the unique capability of retrieving reliable 4D dynamics even at unseen time points using ultra-sparse spatiotemporal data, which will open up new possibilities for 4D X-ray image reconstruction of fast dynamics. 
Besides our XMPI demonstration, we envision that other X-ray imaging modalities can also benefit from our physics-informed workflow, such as time-resolved tomography at modern large-scale X-ray facilities and X-ray laboratory sources. 
The paper is structured as follows. 
First, we introduce the XMPI configuration, which frames the simulation and the 4D reconstruction of this work. 
Second, we introduce the simulation of the droplet collision process under XMPI configuration as a showcase for our proposed reconstruction method.
Third, we describe our proposed physics-informed 4D reconstruction method based on a self-supervised deep learning scheme.   
Fourth, we demonstrate the reconstruction results using 4D-PIONIX and validate its unique capability by comparing it with 4D-ONIX, the state-of-the-art 4D reconstruction method for XMPI.
Finally, we conclude with an outlook for possible future applications and developments.\\

\normalsize

\section{XMPI configuration}
\begin{figure}[!ht]
    \centering
    \includegraphics[width=0.6\textwidth]{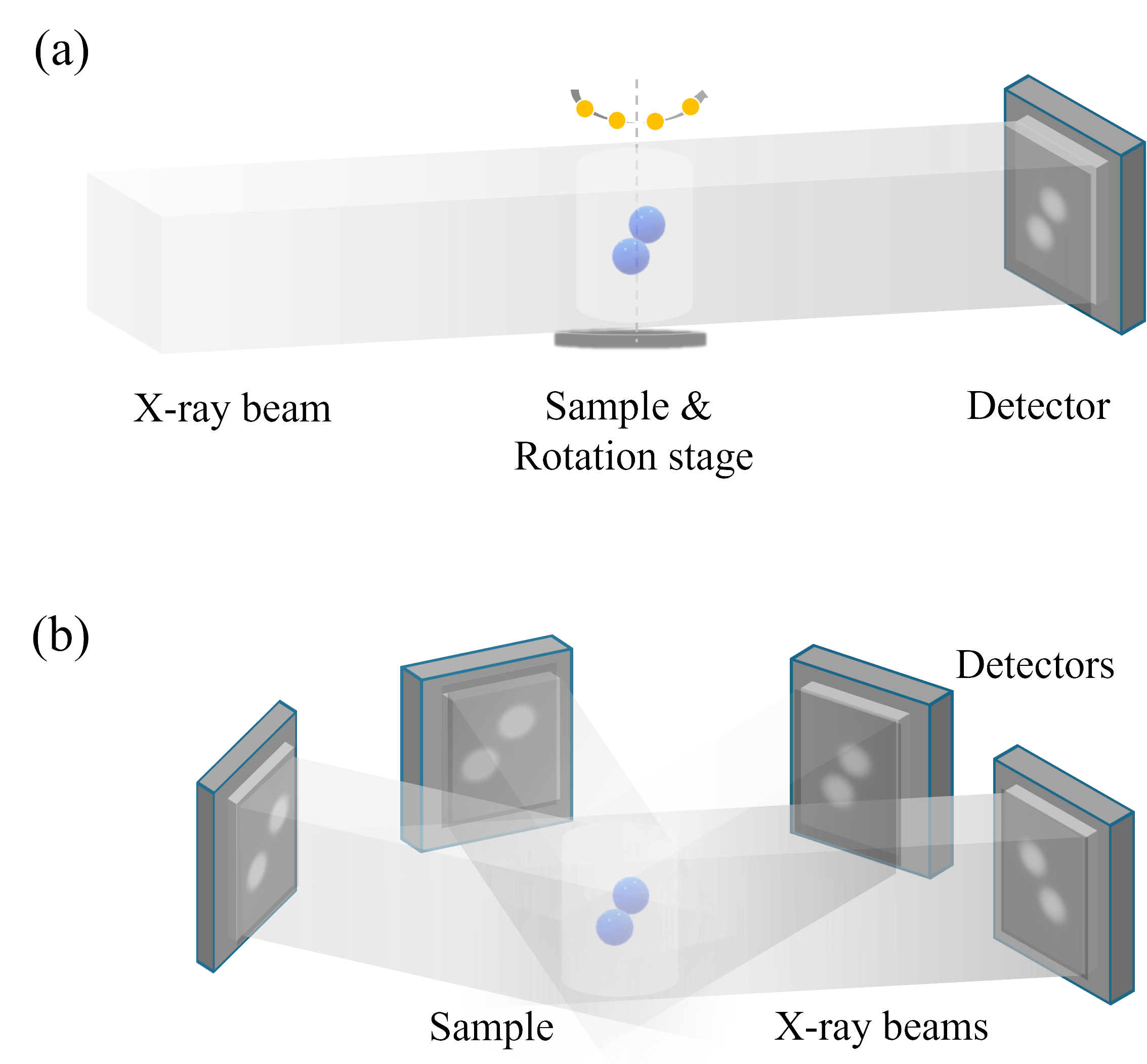}
    \caption{Conceptual configuration of time-resolved X-ray tomography and XMPI. (a) Time-resolved X-ray tomography requires continuous rotation of the sample to acquire projection images from different angles over time; (b) XMPI is a rotation-free technique that generates multiple secondary X-ray beams that illuminate the sample from different angles simultaneously. This figure is adapted from Ref.~\cite{zhang2024phd}.}
    \label{fig:tomo_vs_xmpi}
\end{figure}

Figure~\ref{fig:tomo_vs_xmpi} depicts the conceptual configuration of the traditional time-resolved X-ray tomography setup and the XMPI setup. 
On one hand, traditional time-resolved X-ray tomography requires continuous rotation at a speed suitable to track the studied process.
Such a rotation speed can potentially alter the observed dynamics because of the induced centrifugal force. 
For example, a 500 g-force can be exerted at an acquisition rate of 1000 tomograms per second assuming a $1~\text{mm}$ radius~\cite{Garcia-Moreno2021_tomoscopy}. 
On the other hand, XMPI is a 4D rotation-free X-ray imaging technique, which is more suitable for capturing rotation-sensitive dynamics with high temporal resolution~\cite{xmpiXFEL,xmpiESRF}. 
XMPI relies on high-brilliance X-ray sources and multiple crystal beam splitters~\cite{bellucci2024development} to split the primary X-ray beam into several secondary beams that allow the simultaneous illumination of a sample from different angles. 
A set of synchronized ultra-fast detectors (beyond kHz) is needed to record projection images from different angles.
One can perform 4D reconstruction in the follow-up data processing step.
It is important to note that the state-of-the-art XMPI setup contains less than ten projections when aiming at kHz dynamics and beyond. 
Moreover, due to the recording length limit of some fast cameras, such as Shimadzu Hyper Vision CMOS camera (HPV-X2)~\cite{vagovivc2019megahertz}, the number of usable recorded images for reconstruction is usually limited.
Therefore, the XMPI dataset for 4D reconstruction can be ultra-sparse in terms of both projection angles and time points.
Although the approach presented here can be applied to less restrictive approaches such as time-resolved tomography, in the context of this work, we demonstrate the 4D reconstruction using the simulated data from the XMPI experiment, one of the most demanding existing scenarios in terms of spatiotemporal conditions. \\

\section{Simulation}
To provide proper datasets for testing our proposed physics-informed reconstruction method, we conducted a numerical simulation of binary droplet collisions. 
The simulation in this work includes two steps, as explained in subsections~\ref{4D_droplet} and ~\ref{X-ray}. 
In the first step, we simulated the 4D dynamics of the collision by numerically solving the partial differential equation that governs such dynamics. 
This step can provide the ground truth for the evaluation of 4D reconstruction. 
In the second step, we simulated the XMPI experiment to acquire 2D projection images at each time point, which are later used for 4D reconstruction.\\  

\subsection{4D Simulation of binary water droplet collisions}\label{4D_droplet}
\newcommand{\vecu}{\bm{u}}
\newcommand{\veczero}{\mathbf{0}}
\noindent
 We simulated the 4D droplet collision process in which two identical droplets with a diameter of 80 \textmu m collide head-on at a constant speed.
This process is governed by the following non-dimensionalized Navier-Stokes equation for the incompressible fluid with potential surface tension $\eta\nabla\psi$ ($\eta$ denotes the chemical potential as defined in Cahn-Hilliard equation~\cite{lovric2019low}):
\noindent
\begin{alignat}{3}
\label{eq:ns1}
\rho(\psi) \big (\partial_t \vecu + \vecu \cdot \nabla \vecu\big) - \frac{\mu(\psi)}{\text{Re}} \nabla \cdot \nabla \vecu  &+ \nabla p + \frac{\eta\nabla\psi
}{\text{We}} = \veczero~\\
\nabla \cdot \vecu &= 0~
\label{eq:ns2}
\end{alignat}
In the equations above, $\psi \in [-1,1]$ is the phase variable, with $\psi =1$ representing pure water, $\psi = -1$ representing pure air, and $\psi \in (-1,1)$ representing a combination phase of water and air; \text{Re} and \text{We} denote non-dimensional Reynolds number and Weber number, respectively; $\vecu$ and $p$ denote the vectorial velocity and scalar pressure fields, respectively; the densities ($\rho$) and the viscosities ($\mu$) are expressed as a function of $\psi$: 
\noindent
\begin{alignat}{3}
\rho(\psi) = \frac{1}{2} \big ( (1 + \psi)\rho_1 + (1-\psi) \rho_2 \big ) \quad \text{ and } \quad
\mu(\psi) = \frac{1}{2} \big ( (1+\psi)\mu_1 + (1-\psi)\mu_2\big )~
\end{alignat}
\noindent\\
We used the open-source framework DUNE~\cite{bastian2021dune,dedner2021extendible} to numerically solve the phase variable $\psi$ and the field variables $\vecu$ and $p$. 
For the simulation carried out in this work, the following parameters were used: $\text{Re} = 200$, $\text{We} = 6.94$, $\rho_1 = 1000\, \text{kg/m}^3$, $\rho_2 = 1 \, \text{kg/m}^3$, $\mu_1 = 10^{-3}\, \text{Ns/m}^2$ and $\mu_2 = 10^{-5}\, \text{Ns/m}^2$.
The 4D simulation contains 75 time points or frames in total, and the time difference between two adjacent frames is 0.075 \textmu s. 
The 3D objects at different time points illustrating different stages of the droplet collision process are shown in the first row of Fig.~\ref{fig:sim}. 

\subsection{Simulation of XMPI datasets}\label{X-ray}
\begin{figure}
    \centering
    \includegraphics[width=0.98\textwidth]{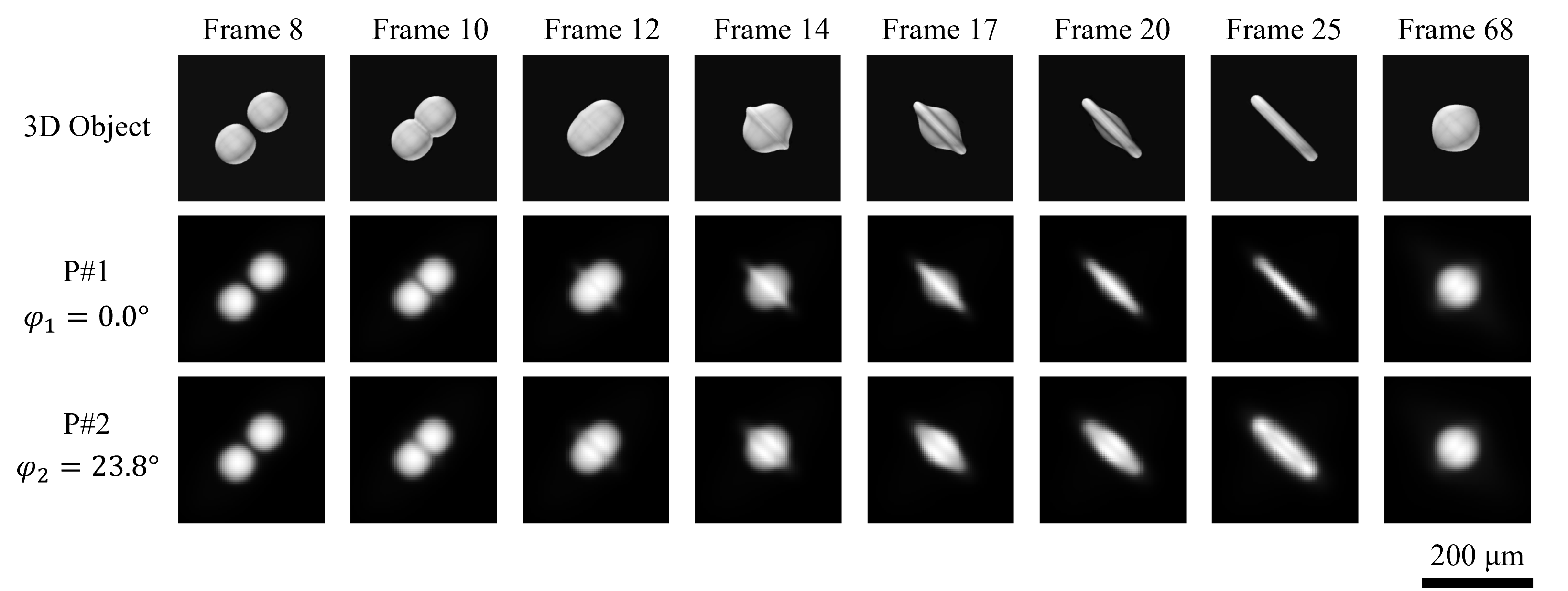}
    \caption{Examples of simulated 3D objects and projection images at eight different time points. The first row shows the simulated 3D object. The second and the third rows show the simulated 2D projection images at $\varphi_1 = \ang{0}$ and $\varphi_2 = \ang{23.8}$, respectively.}
    \label{fig:sim}
\end{figure}

Based on the 4D simulation presented in section~\ref{4D_droplet}, projection images with a pixel size of 4 \textmu m were generated using the X-ray projection approximation (weak scattering)~\cite{paganin2006coherent}. 
We mimicked the challenging conditions of existing MHz XMPI experiments conducted in European XFEL with only two projections per time point~\cite{xmpiXFEL}.
Specifically, we set the X-ray energy as 10 keV and the angle between two projections as $\Delta \varphi = \varphi_2 - \varphi_1 = \ang{23.8}$.
For brevity, we refer to the acquisition of a sequence of projection images from two views as an XMPI experiment. 
Simulated projections at specific time points are depicted in the second and third row of Fig.~\ref{fig:sim}.\\

\noindent
In order to analyze the capability of our proposed physics-informed reconstruction method, we established two datasets with different sparsity in terms of time points based on the same XMPI experiment. 
 The first dataset (75-frame dataset) contains projection images for all 75 time points, corresponding to a frame rate of 13.3 MHz; while the second dataset (15-frame dataset) is a subset of the first dataset where we took the first dataset at a stride of 5 time points, corresponding to a frame rate of 2.7 MHz. \\


\section{4D reconstruction method}
\subsection{4D-PIONIX}
\begin{figure}
    \centering
    \includegraphics[width=0.9\textwidth]{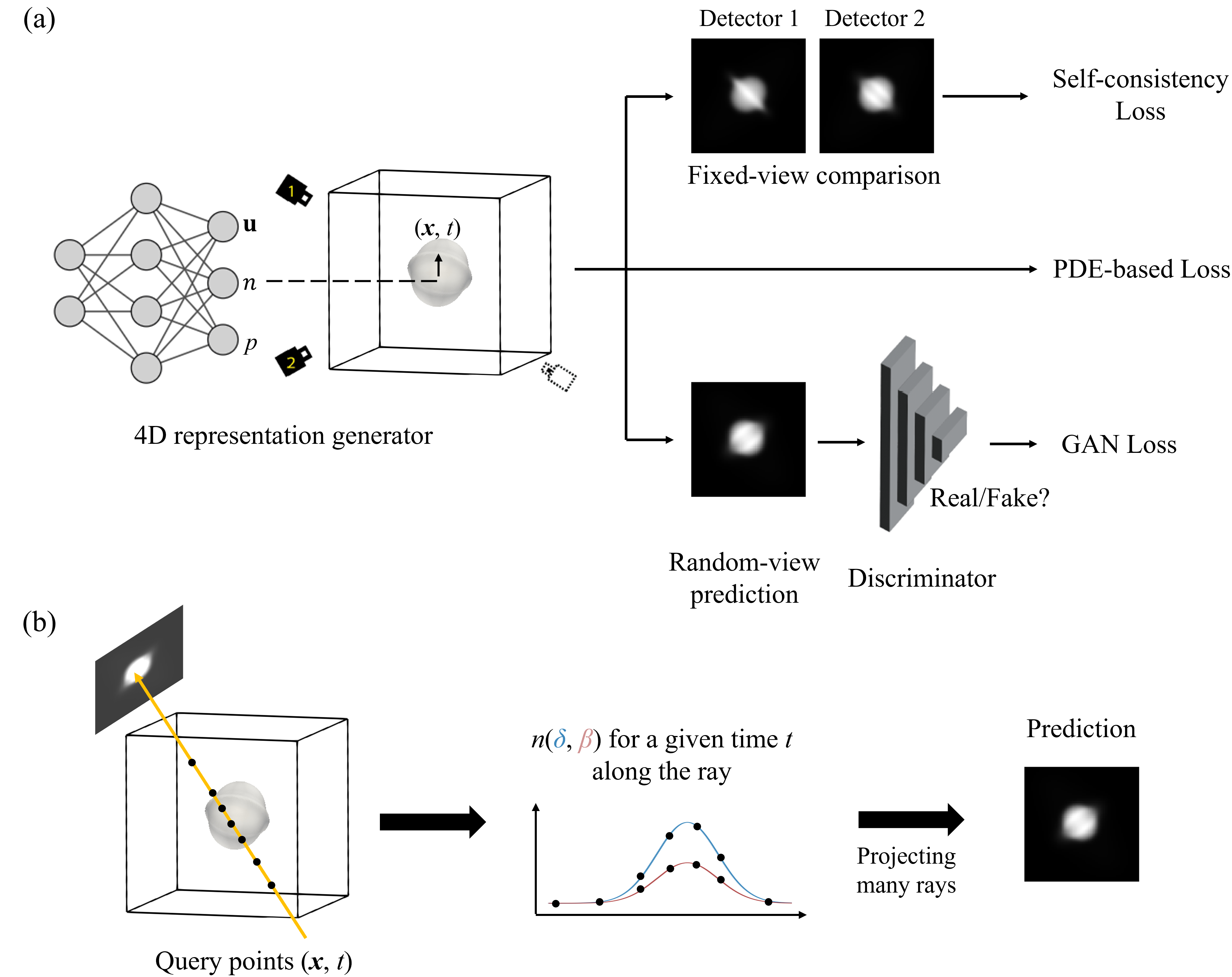}
    \caption{(a) Overview of the 4D-PIONIX workflow. The 4D representation generates the mapping from the 4D spatial-temporal coordinates (\textit{\textbf{x}},\textit{t}) to the physical properties of the sample. The representation is constrained by i) self-consistency between the generated and recorded projections at given angles, ii) the PDE provided by the full physical model, and iii) feedback from the discriminator based on the generated projections at random angles. (b) Generation of the projection images. For each ray directed at a given projection angle, we integrate the refractive index along the ray using the principles of X-ray propagation and interaction with matter. The projection image is formed after sampling all the rays that generate a detector image. This figure is adapted from Ref.~\cite{zhang20244d}.}
    \label{fig:PIONIX_workflow}
\end{figure}

To tackle the challenge of 4D reconstruction from ultra-sparse spatiotemporal data, we designed a physics-informed reconstruction method based on a self-supervised deep learning scheme, as shown in Fig.~\ref{fig:PIONIX_workflow}, inspired by 4D-ONIX~\cite{zhang20244d}, a state-of-the-art algorithm. 
4D-PIONIX combines key concepts of neural implicit representation, generative adversarial neural network (GAN), and physics-informed neural network (PINN) to reliably generate the 4D representation of the observed dynamics.\\

\noindent
Our approach consists of two neural networks: a 4D generator and a discriminator. 
The 4D generator is formed by fully connected multilayer perceptions (MLP) containing five layers of ResBlocks~\cite{he2016deep}. 
It generates the mapping from the 4D spatial-temporal coordinates (\textit{\textbf{x}},\textit{t}) to the physical properties of the sample. 
The physical properties include not only the refractive index of the sample but also two auxiliary variables involved in the full physical model (velocity field $\vecu$ and the pressure field $p$, used to calculate the PDE-based loss function). 
The refractive index is expressed in a complex number with two non-negative components $(\delta,\beta)$, as shown in Eq.~\ref{eq:ior}; its relationship with the phase variable $\psi$ is shown in Eq.~\ref{eq:psi}, where $n_1$ and $n_2$ denote the refractive index of water and air at the given X-ray energy of the XMPI experiment, respectively.
\begin{alignat}{3}
\label{eq:ior}
n &= 1- \delta + i \beta~\\
\label{eq:psi}
n(\psi) = \frac{1}{2} \big ( (&1 + \psi)n_1 + (1-\psi) n_2 \big ) \quad~
\end{alignat}
The refractive index dictates the law of X-ray interaction, allowing us to generate projection images from any angle at any time point based on the 4D representation.
The process of generating projection images~\cite{zhang20244d} is depicted in Fig.~\ref{fig:PIONIX_workflow}(b).
For a given time and a ray directed at a given projection angle, we integrate the refractive index along the ray using the principles of X-ray propagation and interaction with matter under the projection approximation~\cite{paganin2006coherent}.
By assembling all the rays that form a detector image along one direction, projection images are generated.
The discriminator is formed by a convolutional neural network (CNN). The goal of the discriminator is to distinguish the differences between the image patches~\cite{ledig2017photo} from the real (measured) projection images and the predicted projection images by the reconstruction algorithm. 
Using the feedback from the discriminator, the generator can be trained to provide the 4D representation in a higher quality that leads to more indistinguishable projection images by the discriminator.\\

\noindent
Three different losses are involved to constrain the 4D reconstruction and the full physical model: the self-consistency loss, the GAN-loss, and the PDE-based loss. 
Following the implementation of 4D-ONIX, the self-consistency loss and the GAN-loss are described in Eq.~\ref{eq:mse_loss} and Eq.~\ref{eq:gan_loss}, respectively, 
\begin{alignat}{3}
\label{eq:mse_loss}
\mathcal L_\mathrm{MSE} = \sum_{\nu \in \left \{ \alpha, \beta \right \}} &\left \|\mathbf{c}_v - \mathbf{\hat{c}_\nu} \right \|^2_2,\\
{\mathcal L_\mathrm{GAN}} = \mathbb{E}_{\mathbf{c}_v \sim p_D}\log(\mathrm{\mathbf D (\mathbf{c}_v))} &+\mathbb{E}_{\mathbf{\hat{c}_\nu}\sim p_\nu}\log(1-\mathrm{\mathbf D} (\mathbf{\hat{c}_\nu}))
\label{eq:gan_loss}
\end{alignat}
where $\mathbf{c}_v$ and $\mathbf{\hat{c}_\nu}$ denote image patches from the real and generated projections, respectively; $\alpha$ and $\beta$ denote the two angles from which the projections are recorded; $\mathbf D$ denotes the discriminator.\\

\noindent
The PDE-based loss is based on Eqs.~\ref{eq:ns1} and~\ref{eq:ns2}, which can be calculated exploiting automatic differentiation:
\begin{alignat}{3}
\label{eq:pde_loss}
\mathcal L_\mathrm{PDE} = \Big \| \rho(\psi) \big (\partial_t \vecu + \vecu \cdot \nabla \vecu\big) - \frac{\mu(\psi)}{\text{Re}} \nabla \cdot \nabla \vecu  &+ \nabla p + \frac{\eta\nabla\psi
}{\text{We}} \Big\|^2_2 
+\left \| \nabla \cdot \mathbf{u} \right \|^2_2 
\end{alignat}

\subsection{Training details}
In order to efficiently exploit the three loss functions to constrain the 4D representation and avoid the mode collapse problem of GANs~\cite{kodali2017convergence}, we used the following strategy during training. 
In the first stage, for each iteration, we generated the projection images at two fixed angles ($\varphi_1 = \ang{0}$ and $\varphi_2 = \ang{23.8}$) provided in the simulation and optimized the loss function based on $\mathcal L_\mathrm{MSE}$ and $\mathcal L_\mathrm{PDE}$. 
In the second stage, we allowed each iteration to have a 50 percent probability of generating the projection images at random angles, i.e., other than $\varphi_1$ and $\varphi_2$. 
When random angle projections were generated, we optimized the loss function based on $L_\mathrm{GAN}$ and $\mathcal L_\mathrm{PDE}$, as measured projections at such angles are not available.\\

\noindent
We implemented the algorithm in Python 3.9 and PyTorch 1.12. 
We performed the training on NVIDIA V100 GPU with 32 GB of RAM. 
The number of training epochs was adjusted according to the number of frames included in the dataset. 
For example, we trained 9600 epochs for the 15-frame dataset and trained 5400 epochs for the 75-frame dataset. 
In both cases, it took approximately 35 hours to train the model.

\section{Results and discussions}
In this section, we evaluate the performance of 4D-PIONIX and present how it offers a solution to overcome the challenge of 4D reconstruction using an extremely limited dataset from a single simulated XMPI experiment.
In Sect.~\ref{results_1exp}, we evaluate its performance using two datasets corresponding to the same XMPI experiment but a different number of available time points. 
In Sect.~\ref{results_with_4donix}, we compare our approach to 4D-ONIX~\cite{zhang20244d}, a state-of-the-art reconstruction method for sparse-view 4D X-ray imaging. 
In order to retrieve an optimal reconstruction with 4D-ONIX, we used a dataset containing 16 droplet collisions, each of which simulated an XMPI experiment under similar conditions but not identical.
For convenience, we refer to this extra dataset as the 16-experiment dataset. 

\subsection{4D-PIONIX reconstruction using a single experiment}\label{results_1exp}
In this section, we assess the performance of our proposed 4D-PIONIX using the 75-frame dataset and the 15-frame dataset containing only one XMPI experiment, as stated in Sect.~\ref{X-ray}. 
In both cases, we evaluate the quality of the 4D representation at all 75 time points to validate the capabilities of our physics-informed approach. 
Three quantitative metrics are calculated for the evaluation of the reconstruction quality by comparing the reconstruction and the ground truth: the Mean-square Error (MSE); the Dissimilarity Structure Similarity Index Metric (DSSIM)~\cite{wang2004image}, and the estimated resolution based on Fourier Shell Correlation (FSC) with the half-bit threshold criterion~\cite{van2005fourier}. 
For all three quantitative metrics, a smaller number indicates a better reconstruction quality.\\

\noindent
The reconstructed 3D objects based on the 15-frame dataset and the 75-frame dataset using our proposed 4D-PIONIX algorithm are shown in the rows (2) and (3) of Fig.~\ref{fig:results_big_fig}(a), respectively, together with the ground truths in row (1). 
It is important to note that all the 3D ground truths are not accessible to our algorithm, and they are only used for evaluation.
The distributions and the statistics (mean value and the standard deviation) of the 3D metrics over all 75 time points are shown in Fig.~\ref{fig:results_big_fig}(b)-(d) (green and yellow curves) and Table~\ref{tbl:4D-metrics}, respectively.
Besides the 3D metrics, we also calculate the 4D metrics (4D-MSE and 4D-DSSIM) for the entire 3D movie. 
Compared to the 75-frame dataset, the 15-frame dataset leads to a slightly higher 4D-$\mathrm{DSSIM}$ and a slightly lower 4D-$\mathrm{MSE}$.
These quantitative 3D and 4D metrics for the two datasets are comparable, showing the potential of constraining the physical process via the PDE-based loss.\\



\noindent
To sum up, 4D-PIONIX can provide a reliable 4D reconstruction of the entire 75-frame sequence using the dataset containing only a single XMPI experiment. 
Specifically, the quality of the 4D reconstruction using the 15-frame dataset is comparable with the one using the 75-frame dataset, even for the time points that are not available in the dataset. 
It clearly shows the ability of 4D-PIONIX to capture the physical process and reproduce it even at unseen time points.\\

\subsection{Comparison with 4D-ONIX}\label{results_with_4donix}
We have shown that 4D-PIONIX has the potential to reconstruct 4D dynamics using as few as 15 time points from a single XMPI experiment. 
In this section, we verify the unique capability of our proposed 4D-PIONIX method by comparing it with 4D-ONIX, a state-of-the-art 4D reconstruction method.\\

\begin{figure}[!htbp]
    \centering
    \includegraphics[width=0.80\textwidth]{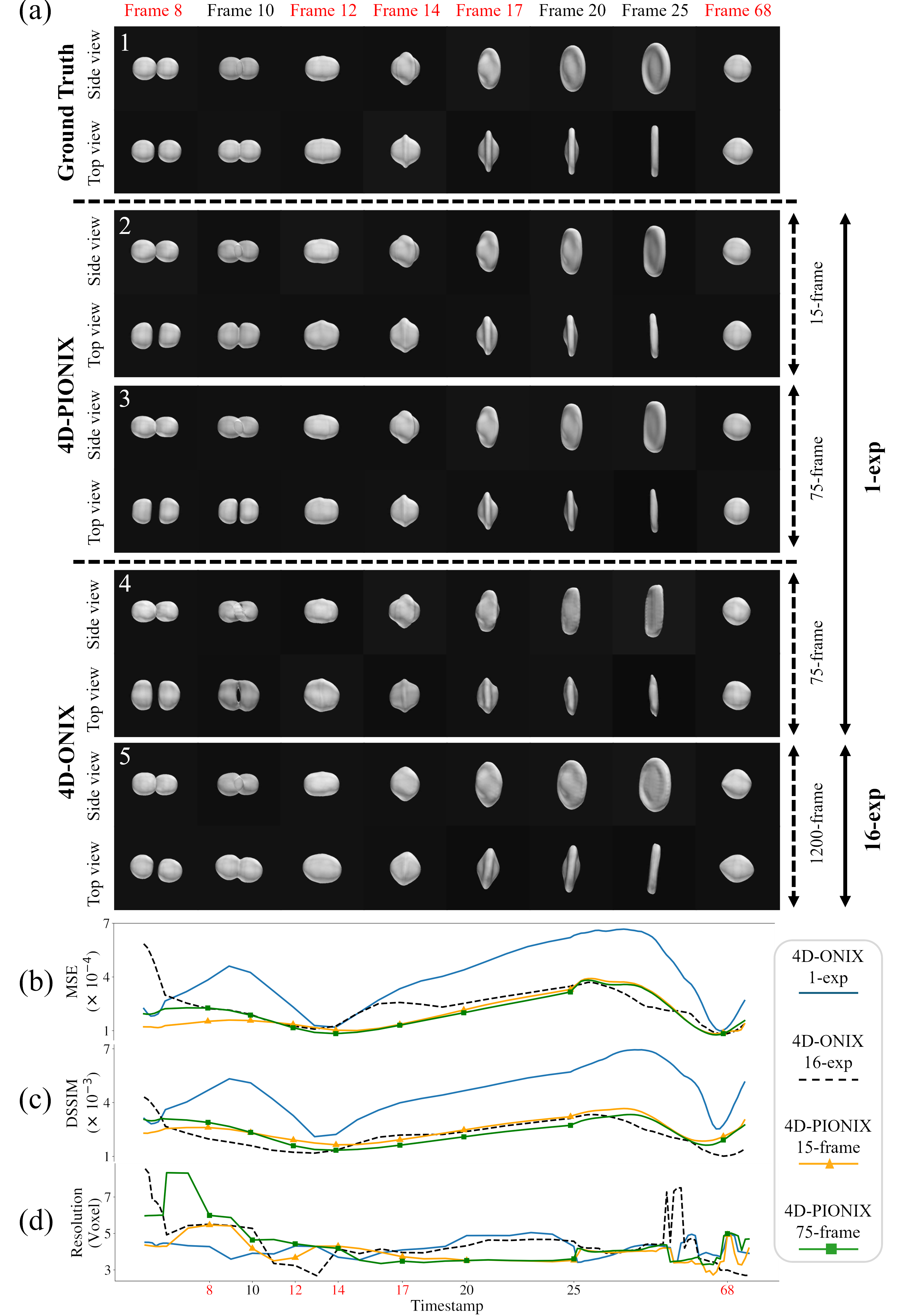}
    \caption{Reconstruction results. (a) Ground truths (1) and reconstructions using 4D-PIONIX, 15-frame dataset (2); 4D-PIONIX, 75-frame dataset (3); 4D-ONIX, 1-experiment dataset (4); 4D-ONIX, 16-experiment dataset (5) at eight time points. At the time points marked in red, projection images are unavailable in the 15-frame dataset, while projection images are available for all four datasets at the time points marked in black. (b)-(d) Comparison of the distribution as a function of time of 3D MSE (b), 3D DSSIM (c), and 3D resolution estimated by FSC analysis (d) under all four reconstruction settings.}
    \label{fig:results_big_fig}
\end{figure}

\noindent
As stated in Ref.~\cite{zhang20244d}, 4D-ONIX works better when more similar XMPI experiments are included in the dataset. 
Hence, in order to ensure that 4D-ONIX works properly, we first implemented 4D-ONIX for both the single-experiment dataset (the same as the 75-frame case in the previous section) and the 16-experiment dataset containing a total of 1200 frames analogously to what is done in Ref.~\cite{zhang20244d}. 
It is important to note that 4D-ONIX utilizes the projection images and the encoder as the input of the neural network. 
Therefore, it cannot provide a proper 4D reconstruction for the unseen time points. 
In other words, reconstructing all 75 time points using the 15-frame dataset as input is not feasible with 4D-ONIX, which highlights the unique capability provided by 4D-PIONIX. 
The reconstructed objects at several time points using 4D-ONIX are shown in the rows  (4) and (5) of  Fig.~\ref{fig:results_big_fig}(a).
These results are consistent with the findings in Ref.~\cite{zhang20244d} that 4D-ONIX prefers several XMPI experiments over the same or similar dynamics seen from different viewpoints to reconstruct 4D processes.\\ 

\begin{table}[bt]
\centering
\caption{4D and 3D metric results for 4D-PIONIX and 4D-ONIX.
}
\label{tbl:4D-metrics}
\vspace{.4cm}%
\begin{tabular}{c c|c c c c}
\hline
&  & \multicolumn{2}{c}{\textbf{4D-PIONIX}} & \multicolumn{2}{c}{\textbf{4D-ONIX}}\\ 
\multirow{2}{*}{\makecell[c]{\textbf{Dataset} \\ \textbf{conditions}}} & {\makecell[c]{\# Experiments}} & 1 & 1 & 1 & 16   \\ 
& {\makecell[c]{\# Time points}} & 15 & 75 & 75 & 1200\\ \hline

\multirow{3}{*}{\makecell[c]{\textbf{3D metrics} \\ \textbf{(mean $\pm$ std)}}} & {3D-MSE $\times 10^{-4}$} & $2.2\pm1.1$ & $2.3\pm1.0$ & $4.2\pm2.0$ & $2.5\pm1.1$\\
& {3D-DSSIM $\times 10^{-3}$} & $2.7\pm0.6$ & $2.5\pm0.6$ & $5.0\pm1.5$ & $2.3\pm0.8$\\
& {\makecell[c]{3D-resolution (voxels)}} & $3.9\pm0.5$ & $4.3\pm1.0$ & $4.2\pm0.4$ & $4.4\pm1.3$ \\ \hline

\multirow{2}{*}{\textbf{4D metrics}} & {4D-MSE $\times 10^{-4}$} & 2.22 & 2.27 & 4.23 & 2.52\\
& {4D-DSSIM $\times 10^{-3}$} & 2.69 & 2.51 & 5.12 & 2.34\\ \hline
\end{tabular}
\end{table}


\noindent
Second, we compare 4D-ONIX results with the 4D-PIONIX results shown in the previous section. 
Table~\ref{tbl:4D-metrics} summarizes the conditions of the dataset, the 4D-metrics of the reconstruction, and the 3D-metrics with respect to time when using 4D-PIONIX and 4D-ONIX, respectively. 
Compared with 4D-ONIX using a 16-experiment dataset, 4D-PIONIX using a single-experiment dataset (either using the 15-frame dataset or the 75-frame dataset) provides similar statistics on the 3D metrics, as well as a slightly lower 4D-MSE and slightly higher 4D-DSSIM.
Furthermore, in order to compare the distributions of the 3D metrics (3D-MSE, 3D-DSSIM, and 3D resolution) with time, Figs.~\ref{fig:results_big_fig}(b)-(d) are given, consolidating the similarity of the 3D metrics shown in Table~\ref{tbl:4D-metrics} among these three reconstruction conditions. 
In short, except for the case when using 4D-ONIX on a single-experiment data, all three reconstruction conditions could retrieve high-quality 4D reconstructions across space and time.\\

\noindent
To sum up, 4D-PIONIX outperforms 4D-ONIX in two key aspects. 
First, 4D-PIONIX using only a single experiment can provide 4D reconstruction with a quality comparable to 4D-ONIX using multiple experiments, indicating its capability of addressing the challenges from ultra-limited data. 
Second, thanks to the full physical model utilized in 4D-PIONIX, it can provide 3D reconstruction at unseen time points and potentially enhance the temporal resolution by capturing the physical process.
Here, we provide a qualitative explanation by comparing the workflow of 4D-PIONIX and 4D-ONIX. 
On the one hand, in 4D-ONIX, the input to the generator includes not only the spatial-temporal coordinates (\textit{\textbf{x}},\textit{t}) but also the encoded version of the projection images from both views. 
The latter, especially the CNN-based encoder, plays an important role during the training process as it helps to transfer knowledge across different similar experiments. 
Hence, 4D-ONIX is essentially a data-driven approach and can only provide proper 3D reconstruction at time points when proper projection images are given. 
On the other hand, in 4D-PIONIX, the input to the generator is simply the spatial-temporal coordinates. 
It means that even at unseen time points, we can still utilize the PDE-based loss $\mathcal L_\mathrm{PDE}$ to constrain the 4D representation.  
Such advantages of 4D-PIONIX significantly reduce the time and difficulty of operating ultrafast 3D imaging experiments like XMPI. 
Moreover, studies of stochastic dynamics can deeply benefit from 4D-PIONIX, as such experiments are typically unreproducible.\\   

\noindent
Our proposed 4D-PIONIX method also faces several challenges.
First, modeling the full physics of the observed dynamics is usually a challenging task, but it is crucial in the 4D-PIONIX workflow. 
Second, although 4D-PIONIX can accept a much smaller dataset compared with 4D-ONIX, the total computation time does not decrease significantly. 
This is due to the heavy computation map when we calculate the PDE-based loss, especially the second-order derivative. 
It is likely that this problem can be solved by implementing numerical differentiation~\cite{chiu2022can} instead of fully relying on automatic differentiation.     

\section{Summary and outlook}
In this work, we present 4D-PIONIX, a fully physics-informed reconstruction tool for 4D X-ray imaging. 
By including a full physical model of the observed dynamics and the physics of X-ray propagation and interaction with matter, 4D-PIONIX can provide reliable 4D reconstruction even using ultra-sparse spatiotemporal data. 
Moreover, it has the potential to provide proper 3D reconstructions by using the full physical model as a constraint, even at time points when recorded projection images are not available. 
We envision that 4D-PIONIX will open up new possibilities for 4D X-ray image reconstruction of fast dynamics in various X-ray imaging modalities. 
For example, regarding time-resolved tomography, the reconstruction workflow combining the full physical model can help reduce the rotation-speed requirements and the number of projections needed for reconstruction, without compromising the spatiotemporal resolution. 
Beyond 4D reconstruction, our physics-informed workflow may also be extended to retrieve unknown physical parameters~\cite{raissi2019physics} of the physics model in the near future, paving the way for investigations of various rapid 4D dynamics, such as fluid dynamics and composite testing. 

\ack{}

We are grateful to Z. Matej for his support and access to the clusters at MAX IV. We acknowledge LUNARC, the Centre for Scientific and Technical Computing at Lund University, for providing computational resources for simulation. 
This work received funding and support from the ERC-2020-STG, 3DX-FLASH (948426), Swedish Research Council, AI-Twin (2024-04904), and the HorizonEIC-2021-PathfinderOpen-01-01, MHz-TOMOSCOPY (101046448).\\

\clearpage
\printbibliography
\end{document}